\documentclass[%
reprint,
superscriptaddress,
amsmath,amssymb,
aps,
]{revtex4-2}

\usepackage{graphicx}% Include figure files
\usepackage{color}
\usepackage[normalem]{ulem}
\usepackage{dcolumn}% Align table columns on decimal point
\usepackage{bm}% bold math

\begin{document}
	
	\title{{\Large \bf{Comments on: ``Optimized thermal control of a dual-wavelength-resonant nonlinear cavity''}}} 
	
	\author{Roman Schnabel}
	\email{roman.schnabel@uni-hamburg.de}
	\affiliation{%
		Institut f\"ur Quantenphysik und Zentrum f\"ur Optische Quantentechnologien\\
		Universit\"at Hamburg, Luruper Chaussee 149, 22761 Hamburg, Germany}

\date{\today}

\begin{abstract} 
This comment addresses the publication [Appl. Opt. 65, 7090 (2026)], \nocite{Meylahn2026}
which targets an alleged ``optimization'' of a patented invention conceived by the author. The invention had introduced a multi-temperature management scheme for a dual-wavelength-resonant nonlinear cavity that optimized the crystal temperature around the beam waist for phase-matching, while utilizing independently adjustable temperatures in other regions of the same crystal to simultaneously satisfy the double-resonance condition. A foundational form of this invention was initially utilized by co-author H. Vahlbruch without the author's prior knowledge during the construction of the squeezed-light laser for the Virgo gravitational wave detector. In their publication, the three authors now propose a variant that reduces the maximum achievable temperature differential. Here, I demonstrate that this approach does not constitute an optimization but, on the contrary, represents a regression that dismantles the invention. In general, a high-quality implementation of this invention fundamentally depends on maximizing the achievable temperature differential. Only then can the problematic phase differences---introduced by dielectric coatings, Gouy phase shifts, and suboptimal poling lengths at the crystal facets---be effectively compensated for, particularly in short crystals. If desired, smaller temperature differences can always be selected dynamically anyway.
\end{abstract}

\maketitle

\noindent
{\bf 1. INTRODUCTION}\\[2mm]
Optical cavities containing a nonlinear medium, such as LiNbO$_{3}$ or periodically poled KTP, are widely utilized for sum- and difference-frequency generation of laser radiation. They form the foundational basis for second-harmonic generators (SHGs), optical parametric oscillators (OPOs), and sources of squeezed vacuum states \cite{Wu1986} (``squeeze lasers'' \cite{Schnabel2022sl}). The efficiencies of these processes enhance drastically when all participating wavelengths are simultaneously resonant, and phase-matching is optimized. Non-critical phase-matching, as well as the fine-tuning in periodically poled crystals, is typically achieved via temperature control. In general, the phase-matching temperature does not coincide with the simultaneous cavity resonance of the involved wavelengths. This discrepancy arises because the impact of the (identical) Gouy phase scales inversely with the respective wavelengths \cite{Lastzka2007}, and because different wavelengths exhibit different penetration depths into the dielectric mirror coatings. Furthermore, the dicing of quasi-phase-matched, periodically poled crystals typically results in uncontrolled thicknesses of the poling domains at the crystal facets, severely limiting the ability to achieve optimal phase-matching under multi-resonance conditions. Fig.\,1 introduces the patented three-temperature management scheme to compensate for these effects \cite{Schnabel2019patent}.
\begin{figure} 
\includegraphics[width=1\linewidth]{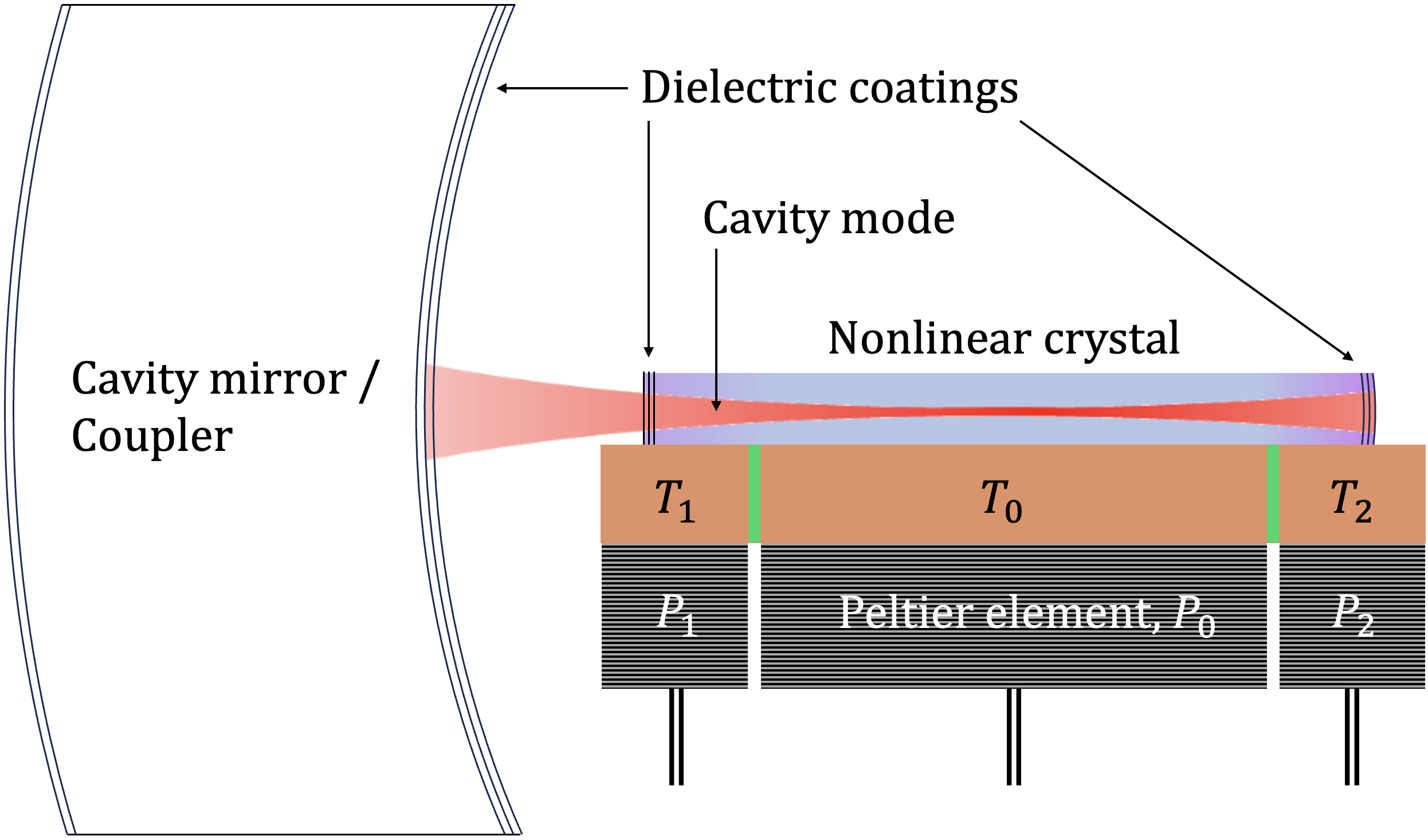}
\caption{
{\bf Multiple-temperature management} --- Schematic of a semi-monolithic (hemilithic) standing-wave cavity for laser light frequency conversion and squeezed vacuum generation. In non-critically phase-matched crystals such as LiNbO$_{3}$ and quasi-phase-matched crystals like periodically poled KTP (PPKTP), fine-tuning of the phase-matching condition is achieved via the crystal temperature to maximize conversion efficiency. A control loop adjusts a piezoelectric actuator behind the coupling mirror to tune the air gap, maintaining resonance for one of the laser wavelengths. Typically, the second involved laser wavelength is non-resonant because the three dielectric mirror coatings introduce different optical path lengths, and the impact of the (identical) Gouy phase scales inversely with the respective wavelengths. In periodically poled crystals, an uncontrolled domain length at the crystal facet could lead to destructive interference of the converted radiation after one cavity round trip. The patented invention \cite{Schnabel2019patent} utilizes two or more independently adjustable temperature zones to simultaneously achieve optimal phase-matching at the beam waist (\(T_{0}\)) and dual-wavelength resonance. These temperatures are monitored and stabilized to their target values via control loops using Peltier elements (\(P_{1,2,3}\)). The efficiency and applicability of this invention---particularly for shorter crystals---fundamentally depend on the capability to establish large temperature differentials across the crystal when required.
}
\label{fig:1}
\end{figure}

~\\ \noindent
{\bf 2. MULTIPLE TEMPERATURES}\\[2mm]
Multi-temperature concepts were realized parallel to and independently of each other by Morgan Mitchell's research group and my own. Zielinska et al.\,\cite{Zielinska2017} designed a monolithic KTP cavity that was poled for quasi-phase-matching exclusively in its central region. The advantage of this approach is that even small temperature differences between the unpoled edge zones and the poled crystal center suffice to establish both phase-matching and double resonance. The downside, however, is the necessity of manufacturing a segmented crystal. In contrast, my concept dispenses with segmentation entirely. It is based on the insight that while optimal phase-matching is critical primarily within the beam waist, different temperatures can be sustained in regions where the beam has already expanded, even if the underlying nonlinear material remains identical throughout the crystal.
Fig.\,1 illustrates this scenario for a semi-monolithic standing-wave cavity. Uniform thermal contact with the planar crystal is provided by a flat assembly comprising three separate copper blocks. The doctoral thesis of Axel Schönbeck (2014–2018) was dedicated to testing my core concept. Employing three Peltier elements and three temperature sensors in three independent control loops, he successfully demonstrated simultaneous double resonance at 1550\,nm and 775\,nm, achieving a squeeze factor of 13\,dB at the telecommunications wavelength of 1550\,nm for the first time \cite{Schoenbeck2018phd}. Based on these results, the University of Hamburg filed a patent application for our invention in 2018, which was published in 2019 \cite{Schnabel2019patent}. The patent has since been granted in Europe, Canada, China, and the United States \cite{Schnabel2022patent}.

~\\ \noindent
{\bf 3. MECHANICAL IMPLEMENTATION}\\[2mm]
H. Vahlbruch was aware of our invention and made use of it (without the author's prior knowledge) for the construction of the squeeze laser for the Virgo gravitational wave detector, as documented in Ref.\,\cite{Mehmet2020}. 
However, instead of executing the crystal mounting surface as a uniform planar face, he apparently utilized two separate copper blocks loosely separated by an air gap. This layout complicated the assembly process and likely induced the reported mechanical stress \cite{Meylahn2026}. Consequently, the crystal may have suffered from insufficient contact area with the temperature-controlled base, resulting in mechanical instability.
The publication \cite{Meylahn2026} merely presents one of many conceivable methods to manufacture a single planar surface capable of transferring two distinct temperatures. Their implementation utilizes two metals with differing thermal conductivities that are bolted directly together in mechanical contact. This specific implementation constitutes the entire scope of their publication.
Crucially, the implementation proposed in \cite{Meylahn2026} no longer permits large temperature differentials because it introduces a partial thermal short-circuit. As a result, the patented invention \cite{Schnabel2019patent} is implemented only in a  compromised, weak variant. In contrast, my own implementations rely exclusively on highly thermally conductive copper blocks, which are securely held by a segmented frame made of a heat-insulating material. This architecture ensures excellent stability without any direct contact between the individual copper segments \cite{Hagemann2024phd, Tohermes2025phd}. 

~\\ \noindent
{\bf 4. MISLEADING REPRESENTATIONS}\\[2mm]
Aside from presenting a poorly efficient, purely metallic thermo-mechanical component, the publication offers no novel insights. Instead, it contains several statements prone to creating false impressions. (The following section references refer directly to the commented publication \cite{Meylahn2026}.)
\\
1. Title: The chosen title conceals the fact that multi-temperature control schemes have been previously published \cite{Zielinska2017,Schoenbeck2018phd}. A more accurate and transparent formulation (in line with the authors' actual work) would have been: ``Optimized two-temperature control of a dual-wavelength-resonant nonlinear cavity''.
\\
2. The word ``Optimized'' in the title obviously references the original patented work \cite{Schnabel2019patent}. Regrettably, and contrary to standard scientific practice, this work is not acknowledged in the abstract; rather, it is relegated to an obscure position as Ref.\,[20].
\\
3. Fig.\,2: Subfigure (b) clearly demonstrates that, without the proposed ``optimization'', the crystal exhibits two spatially extended regions with a large temperature difference between them. This represents a significant advantage over the ``optimization'' shown in (a). The partial thermal short-circuit is avoided---a critical point the authors fail to discuss. In scenario (b), the temperature of the second, cooler Peltier element could simply be raised significantly from approximately 24$^\circ$\text{C} to achieve the same double-resonance condition. The discussion surrounding Fig.\,2 obscures the low efficiency inherent to their bimetallic implementation.
\\
4. Fig.\,3: Upon inspecting the figure and reading its caption, the reader is left unaware that the displayed temperature measurements were taken at the bimetallic mount itself, rather than on the crystal. This crucial information is only disclosed in the main text (``The image of the heat sink...''). Due to the thermal conductivity of the crystal, the actual temperature gradient within the crystal will be substantially weaker than what is suggested by the figure. Indeed, the simulation of the intra-crystal temperature gradient in Fig.\,2\,(b) yields a value of only (31\({}^{\circ }\)–27\({}^{\circ }\))/mm along the crystal axis, which must be considered moderate. Consequently, utilizing two copper plates separated by a 1\,mm air gap (b) represents a solid engineering approach rather than a deficient one, contrary to what the authors attempt to imply.
\\
5. Fig.\,4 and Fig.\,5: The authors restrict their analysis to the specific case of an 11.5\,mm long PPKTP crystal mounted on a bimetallic structure consisting of 8.5\,mm copper and 3\,mm steel. They find that this configuration yields a weak temperature gradient of approximately 2\({}^{\circ }\)\(\text{C}/\)mm. This low value is related to the relatively large 3\,mm length of the steel body. Crucially, a more efficient implementation could always be dynamically dialed down in temperature differential to match such a low value if required.
The patented invention \cite{Schnabel2019patent} is designed to operate effectively even with significantly shorter crystals. For instance, we operate monolithic PPKTP crystals with a length of only 2.6\,mm and an identical cross-section of 1\,mm \(\times \) 2\,mm. In such short crystals, the temperature zones are positioned much closer together, and the thermal conductivity of the crystal actively impedes the formation of usable temperature differentials \cite{Tohermes2025,Tohermes2025phd}. The actual engineering challenge in reality is therefore to achieve the {\it highest} possible temperature gradient, not a particularly small one. For crystals whose length is comparable to their height, causing the temperature gradient to remain small, the design concept presented in \cite{Zielinska2017} should prove superior.
\\
6. Fig.\,6 and Conclusion: The data shows that a parametric gain of less than 20 was achieved. This value carries no positive significance, considering that a cavity of the described type can be operated well above threshold, which corresponds to a parametric gain approaching infinity \cite{Yang1994} even without any multi-temperature management. It remains entirely unclear what the authors mean by ``high gain'' in their conclusion. Equally unsupported is their assertion that ``The heating scheme was successfully demonstrated in a double-resonant nonlinear bow-tie resonator at a wavelength of 1064\,nm, achieving high optical parametric gain, which highlights its potential for generating highly squeezed states of light.'' It remains unexplained why no actual squeezing value was presented, and why no value exceeding the established 15\,dB benchmark \cite{Vahlbruch2016} could be demonstrated.
\\[-2mm]

\noindent
{\bf 5. SUMMARY}\\[2mm]
Multi-temperature management schemes for the application discussed herein were pioneered and published in 2017 and 2019 by Refs.\,\cite{Zielinska2017,Schnabel2019patent}. The publication under comment proposes a two-temperature implementation that places two metals with differing thermal conductivities in direct mechanical contact, rather than securing two highly conductive copper elements within a split, thermally insulating frame. 
By tolerating a partial thermal short-circuit, this approach inherently results in poor efficiency.
The authors argue that a reduced temperature gradient is advantageous. In doing so, they overlook the fact that any more efficient system can simply be dialed down dynamically, and that the intrinsic thermal conductivity of the crystal material inherently causes the gradient to be smeared out.
They demonstrate that for a PPKTP crystal with dimensions of 1\,mm \(\times \) 2\,mm \(\times \) 11.5\,mm, a configuration using an 8.5\,mm copper section bolted directly to a 3\,mm steel section is sufficient to maintain their specific gradient. However, this design effectively sacrifices 3\,mm of the total crystal length for optical frequency conversion. Consequently, this approach is fundamentally inapplicable to crystals shorter than 3\,mm. Short nonlinear squeezing resonators are of critical interest in quantum information processing due to their GHz squeezing bandwidths \cite{Tohermes2025}.
In conclusion, the findings presented in the publication are strictly incremental. The manuscript obscures the fact that the foundational insights and core technological advancements had already been fully established in the prior literature \cite{Zielinska2017,Schnabel2019patent}.

%~\\
%\noindent 
%{\bf Disclosures}\\[1mm]
%The author declares no conflicts of interest.\\

%\bibliography{QRC-bib}

\end{document}